\documentclass[superscriptaddress,longbibliography,aps,prl,reprint,showpacs]{revtex4-1}

\usepackage[english]{babel}
\usepackage[T1]{fontenc}
\usepackage{graphicx}
\usepackage{amsmath}
\usepackage{soul}
\usepackage{natbib}
\usepackage{eqnarray}
\bibstyle{aapmrev4-1}
\begin{document}
\title{Layer specific observation of slow thermal equilibration in ultrathin metallic nanostructures by femtosecond x-ray diffraction}
%\title{Quantitative measurement of the surprisingly slow temperature equilibration in metallic nanolayers excited by femtosecond light pulses}

\author{J.~Pudell}
\affiliation{Institut f\"ur Physik \& Astronomie,
 Universit\"at Potsdam, Karl-Liebknecht-Str.\ 24-25, 14476 Potsdam,
 Germany}
\author{A.~Maznev}
\affiliation{Department of Chemistry, Massachusetts Institute of Technology, Cambridge, Massachusetts 02139, USA}
\author{M.~Herzog}
\affiliation{Institut f\"ur Physik \& Astronomie,
 Universit\"at Potsdam, Karl-Liebknecht-Str.\ 24-25, 14476 Potsdam,
 Germany}

\author{M.~Kronseder}
\affiliation{Institut f\"ur Experimentelle und Angewandte Physik, Universit\"at Regensburg, Germany}

\author{C.~Back}
\affiliation{Institut f\"ur Experimentelle und Angewandte Physik, Universit\"at Regensburg, Germany}
\author{G. Malinowski}
\affiliation{Institut Jean Lamour (UMR CNRS 7198), Universit\'{e} Lorraine, Nancy, France}

\author{A.~von~Reppert$^*$}
\affiliation{Institut f\"ur Physik \& Astronomie,
  Universit\"at Potsdam, Karl-Liebknecht-Str.\ 24-25, 14476 Potsdam,
  Germany}
\author{M.~Bargheer} \email{bargheer@uni-potsdam.de}
\homepage{http://www.udkm.physik.uni-potsdam.de} \affiliation{Institut
  f\"ur Physik \& Astronomie, Universit\"at Potsdam,
  Karl-Liebknecht-Str.\ 24-25, 14476 Potsdam, Germany}
\affiliation{Helmholtz Zentrum Berlin, Albert-Einstein-Str.\ 15, 12489
  Berlin, Germany}

\newcommand{\superscript}[1]{\ensuremath{^{\textrm{#1}}}}
\newcommand{\subscript}[1]{\ensuremath{_{\textrm{#1}}}}

\date{\today}
\begin{abstract}
\textbf{
Ultrafast heat transport in nanoscale metal multilayers is of great interest in the context of optically-induced demagnetization, remagnetization and switching. We investigate the structural response and the energy flow in the ultrathin double-layer system Gold (Au) on ferromagnetic Nickel (Ni) by ultrafast x-ray diffraction (UXRD). The penetration depth of light exceeds the bilayer thickness, preventing unambiguous layer-specific information from optical probes. %By UXRD we quantify the ultrafast heat transport at the nanoscale which is highly relevant for optically induced demagnetization and remagnetization. %Although we optically excite the system through the metallic Au film, we find a very rapid heating of the Ni lattice, whereas the heat transfer to the adjacent ultrathin Au lattice is two orders of magnitude slower than predicted by the heat diffusion equation.
Even though the excitation pulse is incident from the Au side, we observe a very rapid heating of the Ni lattice, whereas the Au lattice initially remains cold; the subsequent heat transfer from Ni to the Au lattice is found to be two orders of magnitude slower than predicted by the conventional heat equation and much slower than electron-phonon coupling times in Au. Both observations are independent of the excitation wavelength, although for the same fluence 400\,nm light excites electrons in Au ten times more than 800\,nm light. Simple model calculations show that the different specific heat of electrons in Ni and Au as well as the different electron-phonon coupling rapidly force the majority of thermal energy into the Ni lattice. Our results show that femtosecond UXRD provides an experimental account of heat transport over single digit nanometer distances as the thermal framework for ultrafast spin dynamics. }
\end{abstract}
%\pacs{}
\maketitle
%\subsection{Introduction}
Ultrafast heating and cooling of thin metal films has been studied extensively to elucidate the fundamentals of electron-phonon interactions \cite{wald2016a, lin2008a, hohl2000a,sun1993,heni2016,nico2011a,wang2008} and heat transport at the nanoscale. \cite{cahi2003a, siem2010a, mazn2011a, wang2012a, cahi2014a, choi2014a} The energy flow in metal multilayers following optical excitation attracted particular attention in the context of heat assisted magnetic recording \cite{chal2009a, xu2017} and all-optical magnetic switching. \cite{stan2007a, kiri2010a, mang2014a} The role of temperature in optically induced femtosecond demagnetization is intensely discussed, particularly with regard to multi-pulse switching scenarios. \cite{elha2016a} Two or three temperature models (TTMs) are often used to fit the experimental observations. \cite{koop2010a} The microscopic three-temperature model (M3TM) \cite{koop2010a} which uses Elliot-Yafet spin-flip scattering as the main mechanism for ultrafast demagnetization is often contrasted against super-diffusive spin-transport.  \cite{batt2010a}  Such electron transport is closely related to ultrafast spin-Seebeck effects \cite{sche2014,alek2017}, which require a description with independent majority and minority spin temperatures. The heat flow involving electrons, phonons and spins has been found to play a profound role in ultrafast magnetization dynamics. \cite{choi2015,kiml2017} The description of the observed dynamics in TTMs or the M3TM are challenged by ab initio theory which explicitly holds the non-equilibrium distribution responsible for the very fast photoinduced demagnetization.  \cite{carv2011, carv2013}
The presence of multiple sub-systems (lattice, electrons, and spins), e.\,g.\ in ferromagnetic metals, \cite{heni2016, reid2018} poses a formidable challenge for experimental studies of their coupling and thermal transport on ultrafast time scales when these subsystems are generally not in equilibrium with each other. \cite{carv2011, carv2013, mald2017} Temperature dynamics in metal films are typically monitored using optical probe pulses via time-domain thermoreflectance (TDTR). \cite{cahi2014a} This technique has been a workhorse of nanoscale thermal transport studies, but experiences significant limitations when applied to ultrathin multilayers with individual layer thicknesses falling below the optical skin depth, which are in the focus of ultrafast magnetism research. \cite{choi2015,sche2014,alek2017,esch2013a,khor2014a,esch2014a} Optical probes generally depend on both electronic and lattice temperatures, although in some cases the lattice temperature \cite{choi2014a} or the spin temperature \cite{choi2015} may be deduced. In order to understand the thermal energy flow, it is highly desirable to directly access the temperature of the lattice which provides the largest contribution to the specific heat.
Ultrafast x-ray diffraction is selectively sensitive to the crystal lattice, and  material-specific Bragg angles enable measurements of multiple layers even when they are thinner than the optical skin depth and/or buried below opaque capping layers. \cite{high2007,shay2011a,koc2017a} The expansion of each layer can be measured with high absolute accuracy, in order to determine the amount of deposited heat in metal bilayers that was debated recently. \cite{esch2013a,khor2014a,esch2014a} The great promise of UXRD for nanoscale thermal transport measurements and ultrafast lattice dynamics has already been demonstrated in experiments with synchrotron-based sources. \cite{high2007,sheu2008,shay2011a,koc2017a} However, limited temporal resolution of these experiments ($\sim$100\,ps) only allowed to study heat transport on a relatively slow (nanosecond) time scale and over distances >100\,nm. Ultrafast nanoscale thermal transport research will greatly benefit from femtosecond x-ray sources. While free electron laser facilities are in very high demand, an alternative is offered by laser-based plasma sources of femtosecond x-rays \cite{schi2012a,zamp2009} which lack the coherence and high flux of a free electron laser but are fully adequate for UXRD measurements. \cite{nico2011a,repp2016b,repp2016a} As an example, a recent experiment on 6\,nm thick Au nanotriangles \cite{repp2016b} confirmed the $\tau^0_\text{Au}=5$\,ps electron-phonon equilibration time generally accepted for high fluence excitation of Au. \cite{hohl2000a,hart2004a,delf2006a,nico2011a} For similar fluences ultrafast electron diffraction reported $\tau^0_\text{Ni}=0.75$ to 1\,ps for Ni thin films between room temperature and Curie temperature $T_\text{C}$. \cite{wang2008, wang2010}

In this report, we demonstrate that the use of a femtosecond x-ray probe enables thermal transport measurements over a distance as small as $\sim$5\,nm in a Au/Ni bilayer with thickness $d_\text{Au}= 5.6$\,nm and $d_\text{Ni}= 12.4$\,nm grown on MgO. By monitoring the dynamics of the lattice constants of Au and Ni, we find that the Ni lattice fully expands within about 2\,ps, while the Au lattice initially remains cold even if a significant fraction of the excitation light is absorbed by the electronic subsystem in Au. The Au layer then heats up slowly, reaching the maximum temperature about 80\,ps after optical excitation. The observed thermal relaxation of the bilayer structure is two orders of magnitude slower than the 1\,ps predicted by the heat equation and also much slower than the usual electron-phonon equilibration time $\tau^0_\text{Au}= 1$ to 5\,ps. (see table~1) \cite{hohl2000a,hart2004a,delf2006a} We explain this surprising result in a model (see Fig~1) based on the keen insight into the physics of the thermal transport in Au-Pt bilayers offered in recent studies \cite{wang2012a,choi2014a}, which showed that nonequilibrium between electrons and lattice in Au persists for a much longer time in a bilayer than in a single Au film. We find, furthermore, that on the spatial scale of our experiment thermal transport by phonons in metals can no longer be neglected. Our results underscore challenges for thermal transport modeling on the nanometer scale.  On the other hand, they demonstrate the great potential of the UXRD for monitoring thermal transport under experimental conditions typical for studies of ultrafast magnetism. \cite{roth2012a,koop2010a}

\begin{table*}[t]
\centering
\begin{tabular}{lp{0.5cm}cp{0.5cm}c}
\hline
\hline
Parameter & & Gold & & Nickel  \\
\hline
Lattice specific heat, $C^\text{ph}$ ($10^6\,\text{Jm}^{-3}\text{K}^{-1}$) & & $2.5$ \cite{taka1986a}& & $3.8$ \cite{mesch1981a}\\
Sommerfeld constant, $\gamma^\text{S}$ ($\,\text{Jm}^{-3}\text{K}^{-2}$) & & $67.5$ \cite{lin2008a} & & $1074$ \cite{lin2008a}\\
Electron$-$phonon coupling constant, $g$ ($10^{16}\,\text{W}^{-3}\text{K}^{-1}$) & & $1$ - $4$ \cite{lin2008a} & & $36$ - $105$ \cite{lin2008a}\\
e$-$ph coupling time isolated layers @1000K, $\tau^0$ (ps) & & $1.7$ - $6.7$ & & $1$ - $3$ \\
e$-$ph coupling time equilibrated electrons @1000K, $\tau$ (ps) & & $26$ - $107$ & & $1$ - $3$ \\
%Electron–phonon coupling time, $\gamma$ (ps) & & $0.6$ - $5$ & & $0.$ - $1.5$ \\
Thermal conductivity, $\kappa$ ($\text{Wm}^{-1}\text{K}^{-1}$) & & $318$ \cite{stoj2010} & & $90$ \cite{stoj2010}\\
Thermal conductivity (lattice), $\kappa^\text{ph}$ ($\text{Wm}^{-1}\text{K}^{-1}$) & & $5$ \cite{stoj2010} & & $9.6$ \cite{stoj2010} \\
Expansion coefficient with Poisson correction, $\alpha^\text{uf}$ ($10^{-5}\,\text{K}^{-1}$) & & $3.16$ \cite{nix1941a} & & $2.8$ \cite{nix1941a} \\
\hline
\end{tabular}

\caption{Literature values for material parameters, relevant for modeling the heat transfer after laser excitation. For  C$^\text{ph}$ we use the parameters at room temperature. The e-ph coupling time ranges are calculated for 1000\,K to exemplify the fact that for an equilibrated electron system, the e-ph coupling time in Ni is definitely much shorter (see text).}
\end{table*}

\begin{figure}
  \centering
  \includegraphics[width = 8.7cm]{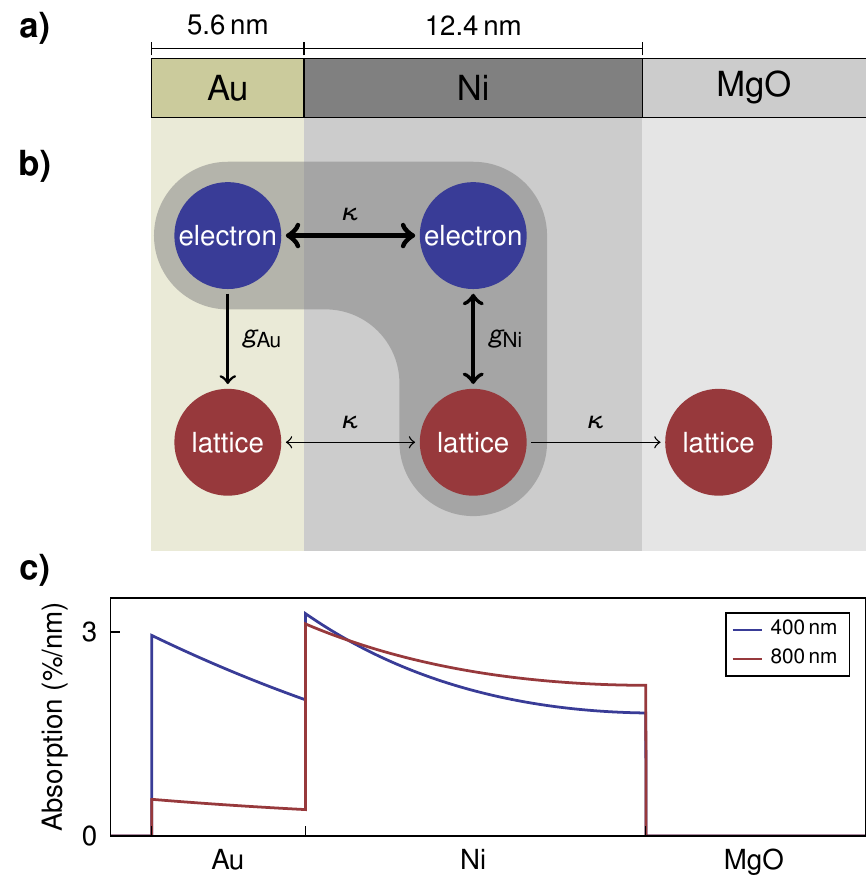}
  \caption{a) Layer stacking of the metallic heterostructure. b) Schematic of the relevant subsystems and their couplings. c) Calculated optical absorption in the metallic thin films.}
\label{fig:Model}
\end{figure}
%In this letter we show direct time-dependent measurements of the lattice temperatures of $d_{Au}=5.6$ nm Au on $d_{Ni}=12.4$ nm Ni on a MgO substrate \cite{footx} by ultrafast x-ray diffraction (see Fig.\ref{fig:Model}a for the structure). We find that irrespective of the excitation wavelength, the Ni film is heated much faster than 1\,ps, whereas the maximum temperature of Au is reached about $\tau=80$ ps after optical excitation. We show that this slow timescale can be essentially rationalized by a model (Fig.\ref{fig:Model}b), where the electron-phonon coupling constant $g_{Au}$ connects the thermal energy of the Au lattice ($\propto C_{Au}^{ph}$) to the combined thermal energy of the Ni lattice and the electron systems of Au and Ni ($\propto C_{com} = C_{Au}^e+C_{Ni}^e+C_{Ni}^{ph} \approx C_{Ni}^{ph}$). These parameters summarized in table 1 determine the surprisingly low rate $1/\tau \approx g_{Au}(\frac{1}{C_{Au}}+\frac{d_{Au}}{d_{Ni}}\frac{1}{C_{Ni}})$ at which energy flows from Ni to Au. Our modeling shows that the heat is transported so slowly into the Au lattice via electronic heat conduction that the lattice heat conduction becomes relevant although in thermal equilibrium the electronic heat conductivity dominates by orders of magnitude.
%We emphasize that the excitation conditions of our experiment are typical for experiments in ultrafast magnetism. \cite{roth2012a,koop2010a}
\begin{figure}[b]
  \centering
  \includegraphics[width = 8.7 cm]{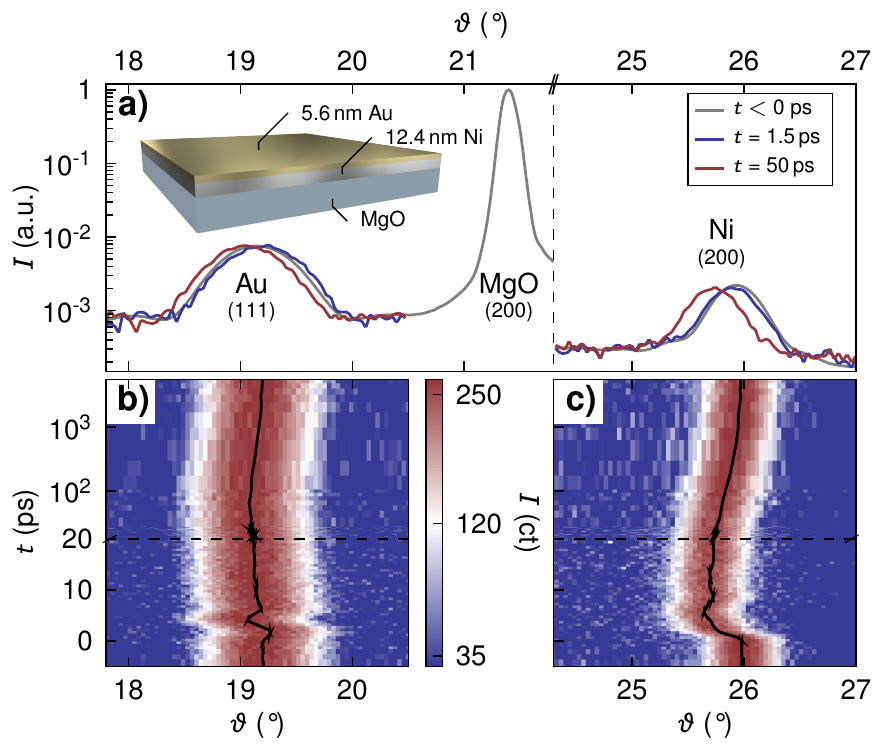}
  \caption{a) X-ray diffraction pattern of the sample (see inset) evidencing the crystalline orientation of the Au and Ni nanolayers. Colored lines visualize transient shifts of the Bragg peaks at selected times. Their full time evolution is shown in panels b) Au and c) Ni along with the respective peak center positions (black line). }
\label{fig:RAW}
\end{figure}
%A central advantage of ultrafast x-ray diffraction UXRD is the material specificity. \cite{koc2017a,repp2016a,repp2016b} %All optical methods inevitably have to deconvolute the signal obtained from the simultaneous absorption and reflection changes in the bilayer.

\section{Experiment and Results}
We use femtosecond laser pulses at 400 and 800\,nm to excite the electron system of Au and Ni through the Au front layer. The sample structure and the calculated absorption profiles are shown in Fig.~\ref{fig:Model}. We note that for 400\,nm pulses the absorbed energy density $\rho^Q_\text{Au,Ni}$ in Au and Ni is similar, whereas for 800\,nm almost no light is absorbed in Au. The much higher absorption of 400\,nm light in Au is a result of the larger real part of the refractive index. \cite{khor2014a,esch2014a} For our 5.6\,nm thick Au film, the destructive interference of light reflected at the interfaces additionally contributes to the suppressed absorption.

%\begin{table}[b]
%\centering
%\begin{tabular}{|l|c|c|c|c|c|c|}
%\hline
% & & & & & &\\%

% &C$^{ph}$ & $\gamma_S $  & g  & $\tau$  &$\kappa$ & $\kappa^{ph}$ \\
% & & & & & &\\
% & ($\mathrm{10^{6}\frac{J}{m^3K}}$) & $(\frac{J}{m^3 K^2})$  & $(10^{16}   \mathrm{\, \frac{W}{m^3 K}}$) & $(ps)$  &$(\frac{W}{m K})$ &$(\frac{W}{m K})$\\
% & & & & & &\\
%\hline & & & &  & &\\
%Gold & 2.5 & 67.5 \cite{lin2008a} & $1 - 4$ \cite{lin2008a}  & $\mathrm{0.6 - 5}$   &318\cite{stoj2010} &5\cite{stoj2010}\\
% & & & & & &\\
%\hline
% & & & & & &\\
%Nickel & 3.8 & 1074 \cite{lin2008a}& $36 - 105 \cite{lin2008a}$ & $0.1 - 1.5$  & 90\cite{stoj2010}&10\cite{stoj2010} \\
% & & & & & & \\
%\hline
%\end{tabular}

The strains $\varepsilon_\text{Au,Ni}$ determined via Bragg's law from UXRD data (Fig.~\ref{fig:RAW}(b,c)) can be converted to lattice temperature changes $\Delta T_\text{Au,Ni}$ and energy density changes $\rho^Q_\text{Au,Ni}$ via
\begin{equation}
 \varepsilon_\text{Au,Ni}= \alpha_\text{Au,Ni}^\text{uf}\,\Delta T_\text{Au,Ni}
 \end{equation}
 \begin{equation}
 \varepsilon_\text{Au,Ni}=\frac{ \alpha_\text{Au,Ni}^\text{uf}}{ C_\text{Au,Ni}}\,\rho^Q_\text{Au,Ni}
 \end{equation}
using effective out-of plane expansion coefficients $\alpha^\text{uf}_\text{Au,Ni}$ and specific heats $ C_\text{Au,Ni}$, which are generally temperature dependent. For our experimental conditions temperature-independent coefficients  are good approximations. The effective expansion coefficients $\alpha^\text{uf}_\text{Au,Ni}$ take into account the crystalline orientation of the films and the fact that on ultrafast (uf) timescales the film can exclusively expand out-of plane, since the uniform heating of a large pump-spot region leads to a one-dimensional situation, as in-plane forces on the atoms by the thermal stresses vanish. For details about $\alpha^\text{uf}_\text{Au,Ni}$ and a description how heat in electrons and phonons drive the transient stress via macroscopic Gr{\"u}neisen coefficients see the methods section.

%
%The lattice constants and strains predicted from equilibrium thermal expansion coefficients require a correction according to the Poisson effect. \cite{lee2008} In cubic materials with (100) surface orientation the ratio of the observed ultrafast (uf) strain and the strain  $\varepsilon^\text{eq}= \alpha^\text{eq}(T) \Delta T$ along the (100) direction calculated from equilibrium value (eq) is $\varepsilon/\varepsilon^\text{eq} = \alpha^\text{uf}(T)/\alpha^\text{eq}(T) = 1+2C_{12}/C_{11}=2.2$ for Ni and would be 2.6 for Au. For the Au (111) cubic crystal surface, the above equation is still valid if the elastic constants are calculated in the rotated coordinate system, in which the x-axis is [111].
%By a transformation of the stiffness tensor of Au to a coordinate system, which is aligned with the (1 1 1) direction
%We find that the newly obtained $C_{11}$ and $C_{12}$ coincidentally yield the same correction factor of 2.2 for Au (111) as for Ni (100).

\begin{figure}
  \centering
  \includegraphics[width = 8.7 cm]{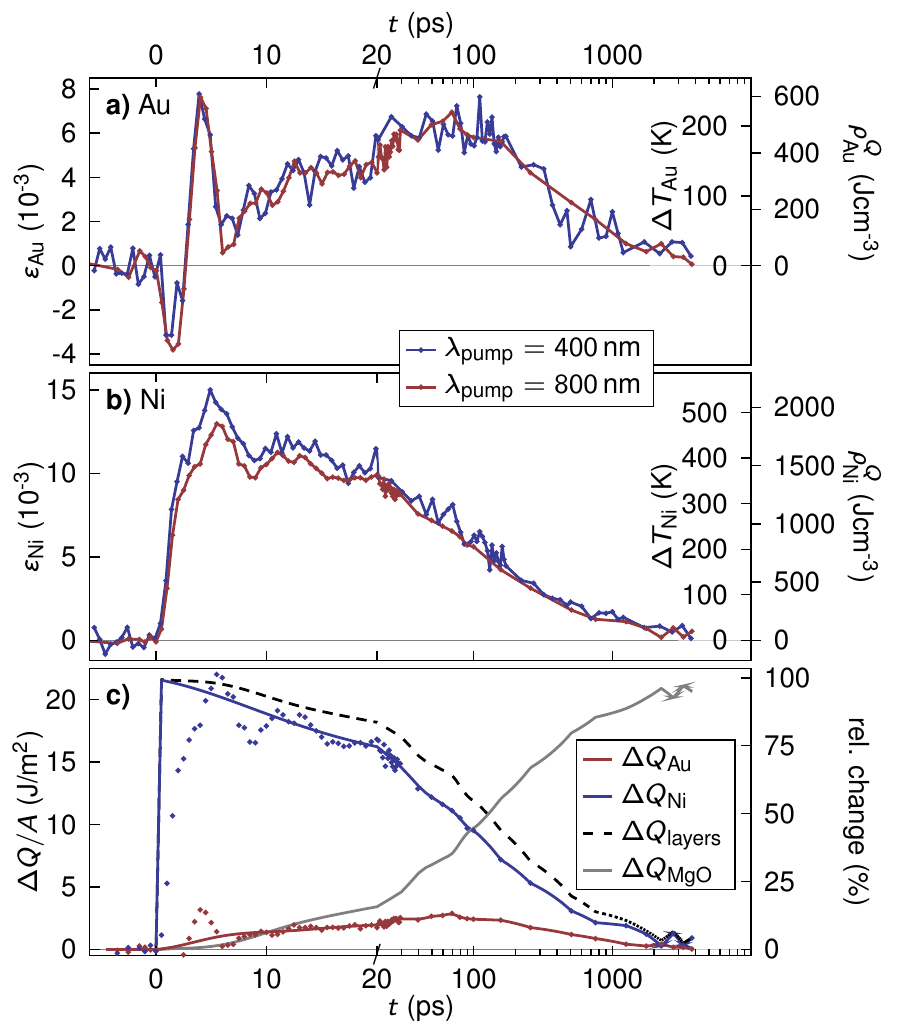}
  \caption{Transient lattice strain $\varepsilon$ in the Au film (panel a) and the Ni film (panel b) as measured by UXRD after excitation with 400\,nm (blue) and 800\,nm (red) light pulses. The right ordinate label the temperature change $\Delta T$ and the energy density $\rho^Q$ calculated from $\varepsilon$. c) Red and blue lines show the energy per unit area $\Delta Q/A$ obtained from panel a) and b) by multiplication with $d_\text{Au,Ni}$. The black dashed line shows the sum of these energies. The grey line is the thermal energy that has been transported into the substrate.   %The right axes indicate the measured energy density $\rho_Q$ according to eq. \ref{eq:Gruen} and the temperature according to eq. \ref{eq:Temp}.
  }
\label{fig:UXRD}
\end{figure}

We now discuss the information that can directly inferred from in the measured transient strains (Fig.~\ref{fig:UXRD}) in the laser-excited metallic bilayer without any advanced modelling. For convenience, we added two right vertical axis to Fig.~\ref{fig:UXRD}a,b) showing the layer-specific temperature and energy density according to eqs.\ (1) and (2). Initially Ni expands, while the Au layer gets compressed by the expansion of the Ni film. Around 3\,ps Au shows a pronounced expansion, when the compression wave turns into an expansion wave upon reflection at the surface. Less pronounced signatures of the strain wave are observed in Ni, as well. A surprisingly long time of about 80\,ps is required to reach the maximum expansion of Au by transport of heat from the adjacent Ni until $T_\text{Au}\approx T_\text{Ni}$. % A simple estimate using the heat equation with the \textbf{thermal conductivity $\kappa$} from table~1 predicts thermal equilibration within less than 1\,ps.
For times $t>100$\,ps, cooling to the substrate dominates the signal.
In Fig.~3c) we show the heat energy $\Delta Q_\text{MgO}$ flowing through a unit area $A$ into the substrate, which we can directly calculate from the measured energy densities via
\begin{multline}
\Delta Q_\text{MgO}(t)/A=%%-(\Delta Q_{Au}(t)+\Delta Q_{Ni}(t))/A=  \\
-d_\text{Au}\Delta \rho^Q_\text{Au}(t)-d_\text{Ni}\Delta \rho^Q_\text{Ni}(t).
\end{multline}
$\Delta \rho^Q_\text{Ni,Au}(t)=\rho^Q_\text{Ni,Au}(t)-\rho^Q_\text{Ni,Au}(0)$ are the changes of the energy densities $\rho_\text{Ni}^Q$ and $\rho_\text{Au}^Q$ with respect to the initially deposited energy densities. Even when the temperatures are equilibrated at $t>100$\,ps, $\rho_\text{Ni}^Q$ and $\rho_\text{Au}^Q$ differ strongly because of the different specific heat of Au and Ni.
Fig.~3c) confirms that within the first 20\,ps the heat energy $\Delta Q_\text{Au}=d_\text{Au}\Delta \rho^Q_\text{Au}$ flowing from Ni into Au is similar to the amount $\Delta Q_\text{MgO}$ transported into the substrate. At about 150\,ps half of the energy deposited in the film has been transported into the substrate. However, leaking a fraction of the thermal energy to the insulating substrate does not explain why the ultrathin Au layer is not much more rapidly heated via electronic heat transport typical of metals.
%\emph{
%As Ni has a larger number of atoms per unit volume than Au, the energy density
%\begin{equation}
%\rho^Q_{Au,Ni}=\int_{T_0}^{T_0+\Delta T} C_{\text{Au,Ni}}(T)\text{d}T.
% \label{eq:Temp}
%\end{equation}
%is larger for Ni in thermal equilibrium, as expressed by the temperature dependent volumetric %specific heats $C_{\text{Au,Ni}}$. It is useful to realize that the temperature dependence of $C_{\text{Au,Ni}}$ and $\alpha_{\text{Au,Ni}}$ is similar, as expressed by the macroscopic Grueneisen-constants $\Gamma \sim \alpha_{\text{Au,Ni}}/C_{\text{Au,Ni}}$, which manifest a linear relation between strain and energy density:   $\varepsilon_{Au,Ni}=\Gamma_{Au,Ni} \rho^Q_{Au,Ni}$. The two right axes of Fig. \ref{fig:UXRD} should emphasizes that the strain is generally linearly related to the stress $\sigma = \Gamma \rho^Q$.}
%However, it turns out to be convenient to use the Grueneisen-constants $\Gamma_{Au,Ni}$, which make the measured variables
%\begin{equation}
%\varepsilon_{Au,Ni}=\Gamma_{Au,Ni} \rho^Q_{Au,Ni}
%\label{eq:Gruen}
% \end{equation}
% proportional to the energy density $\rho^Q_{Au,Ni}$. \cite{koc2017a,repp2016a,repp2016b}

%We have no reason to assume that the non-equilibrium situation in our system strongly favors specific modes with significantly different mode-specific Grueneisen constants. The statistical nature of the process probably leads to an appropriate averaging over many modes rendering eq. \ref{eq:Gruen} a good approximation.

%In order to retrieve a valid lattice temperature, however, we must account for the geometry of the nonequilibrium situation:
 %On the ultrafast timescale the in-plane lattice constant cannot change since expansion waves travel at the sound velocity within about 100 ns across the $500 \mu$m excited nearly homogeneously by the laser. Therefore, the lattice constants predicted from equilibrium thermal expansion coefficients requires a correction according to the Poisson effect. In cubic materials the ratios of the observed ultrafast (uf) strain \cite{foot1} and the strain calculated from equilibrium values (eq) calculated along the (1 0 0) direction is $\varepsilon_{uf}/\varepsilon_{eq} = 1+2C_{12}/C_{11}=2.2$ for Ni and 2.6 for Au. By a transformation of the stiffness tensor of Au to a coordinate system, which is aligned with the (1 1 1) direction we find that the newly obtained $C_{11}$ and $C_{12}$ coincidentally yield the same correction factor of 2.2 for Au(1 1 1) as for Nickel(1 0 0).
 %and Au along (1 1 1), whereas for Au (1 0 0) it reaches up to 2.6. Therefore, we have divided the energy densities $\rho$ and temperatures $T$ displayed in the right vertical axes in Fig. \ref{fig:UXRD} by the factor 2.2. In the hexagonal lattice systems of Dy and Gd the correction factor is only 10$\%$.\cite{repp2016a,koc2017a}

 \section{Modeling}

 Inspired by the recent studies using TDTR \cite{wang2012a,choi2014a} we set up a modified two temperature model graphically represented in Fig.~\ref{fig:Model}b) to rationalize the slow Au heating observed in Fig.~\ref{fig:UXRD}a). % essentially based on arguments brought forward by Cahill.\cite{cahi2003a}
We first justify this simplified modelling.
 The high electron conductivity - potentially including ballistic and superdiffusive electrons - rapidly equilibrates the electron systems of Ni and Au. The fact that the Au layer is equally compressed in the first 2\,ps irrespective of the excitation wavelengths is an experimental proof of the rapid equilibration of electron temperatures. Otherwise the high electron pressure in Au after 400\,nm excitation (cf.\ Fig.~1c)) would counterbalance the compression caused by the Ni expansion. \cite{nico2011a} As Ni has a much larger Sommerfeld constant (table~1) the electronic specific heat $C^\text e=\gamma^\text{S} T$ is dominated by Ni and the ratio of energy densities $\rho_\text{Ni}^Q/\rho_\text{Au}^Q\approx 10$ is large at 1\,ps. %Moreover, the sub-picosecond electron-phonon coupling in Ni dissipates the thermal energy into the Ni phonon system with a large specific heat $C_{Ni}^{ph}$ (Table 1).
 We mention here, that a very large electronic interface resistance \cite{gund2005}, which would prevent a rapid equilibration of electron temperatures in Au and Ni is clearly incompatible with our measurements at 400\,nm. If the electrons would not equilibrate much faster than 1\,ps and effectively remove the heat deposited in the electron system of Au, we would not observe the same strong compression of the Au lattice, since electronic pressure would instantaneously force the Au to expand. \cite{nico2011a,wang2008,wang2010,repp2016a} \footnote{In the diffuse-mismatch model, the electronic interface conductance of metals increases linearly with the temperature and can be calculated from the Sommerfeld constant and the Fermi velocity. \cite{gund2005} Immediately after excitation, the electron temperature reaches several thousand Kelvin, which leads to a sub-picosecond thermalization of the electrons in simulations including the interface resistance.}

  The electron-phonon coupling \emph{constant} in Ni is much larger than in Au (table~1). Consequently, %\cite{nix1941a,Mesc1981a} The electron-phonon coupling in Au is considerably slower and hence
  nearly all photon energy initially absorbed in the electronic system is funneled into the Ni lattice,even when half of the energy is initially deposited in the electronic system of Au as with 400 nm excitation.%, since the hot electrons rapidly equilibrate among Ni and Au.
The electron-phonon coupling \emph{times} $\tau_\text{Au,Ni}^0=C_\text{Au,Ni}^\text{e}/g_\text{Au,Ni}$ for Au and Ni are not very different if the films are not in contact, because the large electronic specific heat $C_\text{Ni}^\text{e}$ of Ni cancels its large electron-phonon coupling constant $g_\text{Ni}$ (see table~1). However, in the bilayer, the electrons in Au and Ni rapidly form an equilibrated heat bath with $C_\text{tot}^\text{e} \approx C_\text{Ni}^\text{e}$. Now only the electron-phonon coupling \emph{constant} determines the coupling time: $\tau_\text{Ni}= C_\text{tot}^\text{e}/g_\text{Ni} \ll C_\text{tot}^\text{e}/g_\text{Au} = \tau_\text{Au}$.

 We start the numerical modeling when a quasi-equilibrium temperature in the combined system $C_\text{com} = C_\text{Au}^\text e+C_\text{Ni}^\text e+C_\text{Ni}^\text{ph}\approx C_\text{Ni}^\text e+C_\text{Ni}^\text{ph} \approx  C_\text{Ni}$ is established after electron-phonon equilibration in Ni around $\tau_\text{Ni}= C_\text{tot}^e/g_\text{Ni} \approx C_\text{Ni}^e/g_\text{Ni} \approx 1$\,ps. Since $C_\text{Ni}^\text{ph}\gg C_\text{Ni}^\text e \gg C_\text{Au}^\text e$ and $d_\text{Ni}>d_\text{Au}$, we refer to the combined system as $C_\text{Ni}$ in the equations. Since the energy stored in each layer is proportional to their thickness and the energy transfer rate from electrons to phonons in Au is proportional to the Au volume $V_\text{Au}\propto d_\text{Au}$, the differential equations describing this special two-temperature model (TTM) represented in Fig.~\ref{fig:Model}b) read
\begin{equation}
d_\text{Au}C_\text{Au}^\text{ph}\frac{\partial T_\text{Au}^\text{ph}}{\partial t}= d_\text{Au}g_\text{Au}(T_\text{Ni}-T_\text{Au}^\text{ph})
\end{equation}
\begin{equation}
d_\text{Ni}C_\text{Ni}\frac{\partial T_\text{Ni}}{\partial t}= d_\text{Au}g_\text{Au}(T_\text{Au}^\text{ph}-T_\text{Ni})
\end{equation}
Note that the two temperatures in this model are the temperature of the Au lattice, $T_\text{Au}^\text{ph}$ and the temperature of the combined system, which is denoted as $T_\text{Ni}$ although this Ni temperature equals the Au electron temperature.
For small temperature changes over which the specific heats are approximately constant, the solution to this system of equations is an exponential decay of $T_\text{Ni}\sim e^{-t/\tau}$ and a concomitant rise of the Au lattice temperature $T_\text{Au}\sim(1-e^{-t/\tau})$ on the characteristic timescale
\begin{equation}
\tau=\frac{1}{g_\text{Au}(\frac{1}{C_\text{Au}}+\frac{d_\text{Au}}{d_\text{Ni}}\frac{1}{C_\text{Ni}})}.
\label{eq:tau}
\end{equation}
%The calculated temperatures are illustrated in Fig. \ref{fig:Theo}a) in order to emphasize that
Due to the small film thickness and the rapid electronic heat diffusion, we do not assume any gradient in the temperatures of each film. At about 1\,ps after excitation we define the initial conditions as $T_\text{Ni}(1\,\text{ps}) = T_\text{Ni}^\text i$ and $T_\text{Au}^\text i\approx 0$. The final temperature after equilibrating the temperatures of the two thin films, neglecting heat transport to the substrate is
\begin{equation}
T^\text f=T^\text i\frac{d_\text{Ni}C_\text{Ni}}{d_\text{Au}C_\text{Au}^\text{ph}+d_\text{Ni}C_\text{Ni}}
\end{equation}
%Here again, $C_{Ni}$ includes the electronic contributions of both films.

\begin{figure}
  \centering
  \includegraphics[width = 8.7 cm]{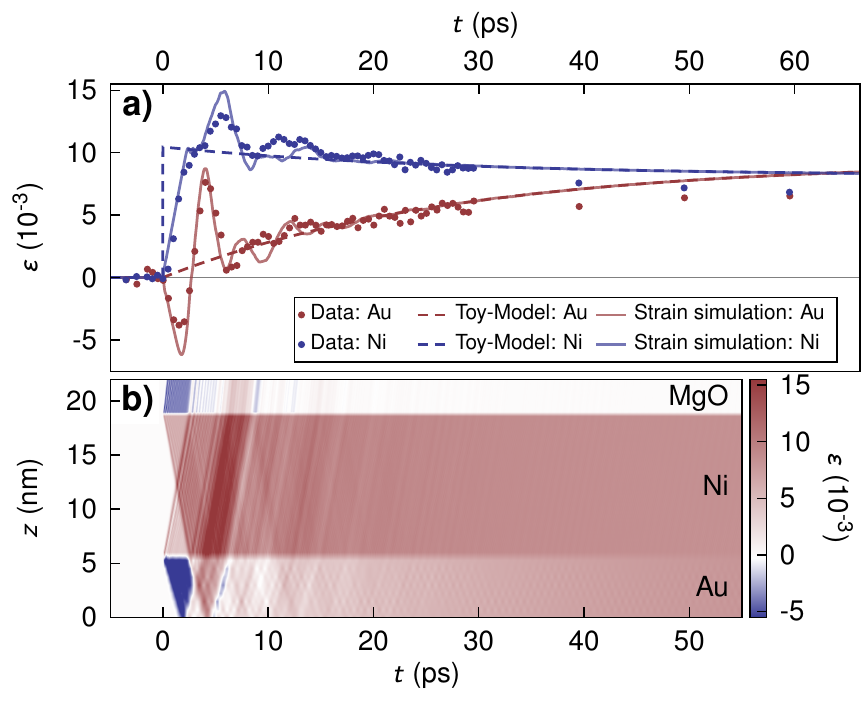}
  \caption{a) Comparison of  models with the experimental data. Dots indicate the measured strain $\varepsilon$. The dashed lines represent the strain calculated from the average heating of the layers according to the model visualized in Fig.~1b). Solid lines are simulations, which are based on this and additionally include the strain waves triggered by the impulsive excitation (see methods section). Heat transport to the substrate is not included. b) Color-coded strain $\varepsilon$ as a function of sample depth and time $t$, which is simulated assuming a spatially homogeneous transient thermal stress in each layer which is proportional to the dashed lines in panel a). Spatial averaging of  the strain $\varepsilon(t)$ in each layer yields the solid lines in panel a). }
\label{fig:Theo}
\end{figure}

 This very simple model (dashed lines of Fig.~\ref{fig:Theo}a)) for the transient quasi-equilibrium temperatures agrees very well with the data. In particular, the exponential rise of $T_\text{Au}$ and the exponential decay of $T_\text{Ni}$ converge around 80\,ps. Deviations at longer times originate mainly from heat transport into the MgO substrate, which is not included in the model (dashed lines).
 %If we add heat transport into the substrate to the simulation, we can achieve agreement also on timescales beyond 100 ps. Such details of the heat transport analysis shall be discussed in a separate publication.

%Does this plot contain a scaling of the absorbed fluence according to the different Bragg angles????

The only fit parameters of our model are the initial temperature $T^i$   and the electron-phonon coupling constant of Au. With our simple model we get the best fit using $g_\text{Au}=6.5\cdot10^{16}$\,W/(m$^3$K), which is somewhat larger than the range from 1 to $4\cdot10^{16}$\,W/(m$^3$K) reported in the literature. \cite{lin2008a,hohl2000a} %In the recent description of energy flow in Aluminium by a nothermal lattice model (NLM) it was shown that electon-phono \cite{wald2016a},  	arXiv:1708.01470 [cond-mat.mtrl-sci]
If - as an example - we reduce the electron-phonon coupling constant to the value of $4\cdot10^{16}$\,W/(m$^3$K), the calculated equilibration of $T_\text{Au}$ and $T_\text{Ni}$ is much too slow. Including electronic interface resistance would make it even slower.%However, if we include phonon heat transport according to the heat equation
%\begin{equation}
%\frac{\partial T}{\partial t} = \alpha \frac{\partial^2 T}{\partial x^2}
%\label{eq:heat}
%\end{equation}
%with a reasonable\cite{??} phononic thermal diffusivity $\alpha = \kappa/C$ into the model, we can recover an excellent fit of the data. Here $C$ is the volumetric specific heat of the metal. Assessing the phononic heat conductivity of metals is difficult, but the literature value $\kappa^{ph}_{Au}\approx 3$\,W/(m$\cdot$ K) fits appropriately?
The missing energy transfer rate, however, can be easily rationalized by phonon heat conductivity $\kappa^\text{ph}$ in these metal films. If we fully disregarded electronic heat conduction in Au, the literature value for $\kappa^\text{ph}_\text{Au}$ given in table~1 would lead to an equilibration of Au and Ni temperature exclusively via phonons three times faster than we observe. The phonon heat transport is probably much less efficient than this prediction because of additional interface resistances for phonon heat transport and because the mean free path of phonons is on the order of the layer thickness. \cite{mazn2011a,cahi2003a} However, we do not attempt to quantify $\kappa^\text{ph}$ and $g_\text{Au}$ here. We only note qualitatively that to conform to the expected values of electron-phonon coupling in Au, the phonon heat conduction must become important in nanoscale multilayers, even though normally the heat conduction in metals is dominated by electrons ($\kappa\gg \kappa^\text{ph}$ see table~1). Phonon heat transport is not included in our numerical calculations, because in fact the heat diffusion equation is not valid at such small length scales below the phonon mean free path. %The heat transport by phonons on the nanoscale contains ballistic contributions. \cite{mazn2011a,cahi2003a}
%We emphasize once more that calculating heat transport with the full heat conductivity $\kappa$ predicts a heating of Au within 1\,ps.
Similarly, a complex theoretical modelling would be required to simulate the heat transport to the substrate, e.\,g.\ by heat transfer from Ni electrons to MgO phonons at the interface. \cite{soko2017}. Fig.~3c) provides a benchmark of the experimentally determined phonon-heat transport into the substrate.

In conclusion, the modified TTM model (eq.~4+5) captures the essence of heat transport between ultrathin metal films: The electrons in Au and Ni are rapidly equilibrated. This is evidenced by the fact that 400 and 800\,nm excitation both initially only heat Ni, regardless of the energy absorbed in Au. For 400\,nm excitation we showed an unexpected shutteling of heat energy between the layers: The electrons first rapidly transport energy from Au into Ni (e-e equilibration $\ll 1$\,ps) before they transport some of the heat back from the Ni phonons to the Au phonons. Finally the heat flows back through Ni towards the substrate. Heat transport by phonons can account for a fraction of the Au heating. The energy transported from the Ni phonons via Ni and Au electrons into the Au lattice is throttled by the weak electron-phonon coupling in Au.
%In conclusion, we have discussed the surprisingly slow heat transport from a 12 nm thick Ni layer into a 6 nm thick Au layer. We showed that - no matter if the photons are absorbed by electrons in Au or Ni - the optical excitation rapidly heats up the Ni lattice, because the electrons in Ni and Au thermalize to the same quasi-equilibrium temperature and the electron phonon coupling and the electronic specific heat in Ni are much larger than in Au. Although the electrons of Au and Ni share the same temperature, heating of the Au lattice is controlled by electron-phonon coupling constant $g_{Au}$ of Au and the specific heat of Ni and the Au lattice. Heat transport by phonon-phonon interaction across the Ni/Au interface adds to the heating.
We believe that our results will have an important impact on ultrafast studies of the spin-Seebeck Effect, super-diffusive electron transport as well as optical de- and re-magnetization.  Precise measurements of the total heat in the system after few picoseconds will help to determine the actually required laser fluence in ultrafast demagnetization studies which currently diverge by an order of magnitude in the literature. \cite{atxi2010, roth2012a} The lattice is not only discussed as the sink of angular momentum in the ultrafast demagnetization: With its dominant heat capacity the lattice constitutes the heat bath which controls the speed of reordering of the spin systems at high fluence. \cite{koop2010a,roth2012a} Our detailed account of heat flow in Ni after photo-excitation must influence the interpretation of MOKE data, which were fitted in previous studies \cite{koop2010a,jiwa2012a} by using a value for the specific heat of the Ni phonon system which is a factor of two below the Dulong-Petit value.

%We presented a direct experimental proof of the peculiar heat transport in metals at the nanoscale and suggest that simultaneous measurements of electron, spin and lattice heat may yield important information for designing future nanoscale devices.
We have demonstrated the power of UXRD in probing nanoscale heat transport in an ultrathin metallic bilayer system which is relevant to current magnetic recording developments such as heat assisted magnetic recording. To understand the all-optical \cite{xu2017} and helicity dependent \cite{aleb2012a} switching in ferrimagnets and two different timescales observed in the demagnetization of transition metals \cite{roth2012a,koop2010a} or rare earths \cite{frie2015, rett2016}, precise calibration of the lattice temperature is crucial.  %UXRD can measure the temperature of layers thinner than the optical skin depth with high accuracy, including layer-specific sensing of multiple buried layers which are not accessible for optical probes.
We are convinced that the direct access to the lattice, the layer-specific information for layers thinner than the skin depth, the conceptual simplicity of the arguments and the experimental geometry make the paper particularly useful for comparisons to previous \cite{esch2013a,khor2014a,esch2014a,roth2012a,koop2010a} and future work on optical manipulation of spins. %The direct and exclusive probe of the lattice is well suited for cross-checking results obtained by all-optical methods.

\section{Methods}
\begin{small}
\subsection{Sample growth and UXRD}

%The Au/Ni bilayer was grown by molecular beam epitaxy on a MgO (001) substrate  %\cite{footxx}
Ni/Au stacks with different Ni and Au thicknesses were grown by molecular beam epitaxy onto a MgO(001) substrate at 100$^{\circ}$C. The MgO(001) substrates were degassed at 350$^{\circ}$C for 10 minutes. The pressure during growth never exceeded $6^{-10}$\,mbar.
We measured the layer thicknesses $d_\text{Au}= 5.6$\,nm and $d_\text{Ni}= 12.4$\,nm of the investigated sample by x-ray reflectivity.%
The 24 lattice planes of Au yield a symmetric (111) Bragg reflection (Fig.~\ref{fig:RAW}a) at $\vartheta=19.29^\circ$, well separated from the symmetric (200) Ni peak at $25.92^\circ$ originating from 70 lattice planes. The lattice strains $\varepsilon_\text{Ni,Au}(t)=-\text{cot}(\vartheta(t))\Delta\vartheta(t)$ perpendicular to the sample surface are directly retrieved from the time-resolved Bragg peak positions $\vartheta(t)$ (Fig.~\ref{fig:RAW}(b,c)). \cite{schi2012a,repp2016a,repp2016b} These ultrafast x-ray diffraction (UXRD) data were recorded at our laser driven plasma x-ray source at the University of Potsdam, that emits 200\,fs x-ray pulses with a photon energy of 8\,keV\@.	The sample was excited by p-polarized 400 and 800\,nm laser pulses of about 100\,fs duration with a pulse energy of 0.3\,mJ and a diameter of 1.5\,mm (FWHM) under an angle of 44$^\circ$ (51$^\circ$) with respect to the surface normal for the Ni (Au) reflection. From the incident fluence of 9 (8)\,mJ/cm$^2$  an absorbed fluence of 3 (2.9)\,mJ/cm$^2$ is calculated for our bilayer system using a matrix formalism, which also yields the absorption profiles at 400 and 800 nm excitation shown in Fig.~1c). \cite{yari1988}

\subsection{Correction of the thermal expansion coefficient}

The effective expansion coefficient $\alpha^\text{uf}_\text{Au,Ni}$ valid for heating a thin epitaxial layer is based on the lattice constants and strains predicted from equilibrium thermal expansion coefficients,  corrected according to the Poisson effect. \cite{lee2008} In cubic materials with (100) surface orientation the ratio of the observed ultrafast (uf) strain and the strain  $\varepsilon^\text{eq}= \alpha^\text{eq}(T) \Delta T$ along the (100) direction calculated from equilibrium value (eq) is $\varepsilon/\varepsilon^\text{eq} = \alpha^\text{uf}(T)/\alpha^\text{eq}(T) = 1+2C_{12}/C_{11}=2.2$ for Ni and would be 2.6 for Au. For the Au (111) cubic crystal surface, the above equation is still valid if the elastic constants are calculated in the rotated coordinate system, in which the x-axis is [111].
%By a transformation of the stiffness tensor of Au to a coordinate system, which is aligned with the (1 1 1) direction
We find that the newly obtained $C_{11}$ and $C_{12}$ coincidentally yield the same correction factor of 2.2 for Au (111) as for Ni (100).

%\subsection{Macroscopic Gr{\"u}neisen concept for ultrafast electronic and phononic stress}
\subsection{Strain waves prove ultrafast electron-equilibration}
The pronounced compression and expansion of the Au layer (see Fig.~4a) clearly originates from the laser-induced stress generated in Ni. In order to show that our modified two temperature model (TTM) predicting negligible energy density in Au immediately after the excitation can quantitatively explain the  signal oscillations, we have used the transient temperatures $T_\text{Ni,Au}(t)$ from our TTM as input parameters for a full thermo-elastic simulation using the udkm1Dsim toolbox which are represented as solid lines in Fig.~4a). \cite{schi2012a} %In order to emphasize that this very simple model already accounts very well for most details of the signal, we show the lattice temperatures $T^{ph}_{Ni,Au,MgO}(t)$ which are used in the model as a color-plot in Fig. \ref{fig:Theo}b).  The temperatures are homogeneous through the depth of each layer and the substrate temperature does not change within the model.
For convenience, Fig.~\ref{fig:Theo}b) shows the spatio-temporal strain map from which the solid lines in Fig.~\ref{fig:Theo}a) are calculated by spatial averaging over the layer for each time delay. Multiple reflections of strain waves at the interfaces are strongly damped by transmission to the substrate.

\subsection{Macroscopic Gr{\"u}neisen coefficients}
Several recent ultrafast x-ray diffraction and electron diffraction experiments on thin metal films have highlighted two contributions of electrons and phonons to the transient stress $\sigma$ which drives the observed strain waves. A very useful concept uses the macroscopic Gr{\"u}neisen coefficient $\Gamma^\text e$ and $\Gamma^\text{ph}$, which relate the energy densities $\rho^Q$ to the stress $\sigma = \Gamma \rho$. While in Au the electronic Gr{\"u}neisen constant $\Gamma^\text{e}_\text{Au}=1.5$ is about half of its phonon counterpart $\Gamma^\text{ph}_\text{Au}=3.0$, in Ni $\Gamma^\text{e}_\text{Ni}=1.5$ is only slightly different from $\Gamma^\text{ph}_\text{Ni}=1.7$. \cite{nico2011a,wang2008} For our analysis the distinction of the origin of pressure in Ni is not very relevant, since the redistribution of energy from electrons to phonons only increases the stress by 15$\%$. In Au the electron pressure is negligible in our bilayer system, since due to the large electronic specific heat of Ni and the sub-picosecond equilibration among the electrons, all the energy is accumulated in Ni. The ab initio modelling discussed in connection to the recent UXRD study on Fe points out that both electron-phonon coupling parameters and phonon Gr{\"u}neisen coefficients depend on the phonon mode. \cite{mald2017, heni2016} While in that study the scattering of x-rays from individual phonon modes selected by the scattering geometry may require a mode-specific analysis, we believe that measuring the lattice expansion via a Bragg-peak shift looks at an average response of the lattice to all phonon modes, and hence a mode-averaged analysis is reasonable if there is no selective excitation of modes with extraordinarily different Gr{\"u}neisen coefficients.
\end{small}
\section{Acknowledgement}
We acknowledge the BMBF for the financial support via 05K16IPA and the DFG via BA 2281/8-1.

\end{document}